\documentclass[aps,pra,nofootinbib,amsmath,amssymb,reprint,letterpaper]{revtex4-1}

\usepackage{graphicx}
\usepackage{color}
\usepackage{natbib}
\usepackage{placeins}

\begin{document}

\bibpunct{}{}{,}{s}{}{\textsuperscript{,}}

 \title{Strictly Localized Molecular Orbitals: Non-Orthogonal Local
 Orbitals Built with Chemical Intuition}

\author{Jeffrey Keyes}
\affiliation{Department of Chemistry, Wesleyan University, Middletown,
  CT 06459}

\author{Carlos A. Jim\'enez-Hoyos}
\affiliation{Department of Chemistry, Wesleyan University, Middletown,
  CT 06459}

\begin{abstract}
  We propose the use of an a priori localization method based on separating the
  molecule into fragments to generate a wavefunction suitable for post
  Hartree-Fock methods. A quasi-Newton approach utilizing Thouless rotation is
  developed to optimize the strictly localized molecular orbitals(sLMO). We
  discuss the properties of the sLMO-HF wavefunction and compare it to the tail
  deleted a posteriori localization methods that are usually done in
  preparation of post Hartree-Fock methods. We then discuss the effects of
  choosing different fragmentation patterns and basis sets.
\end{abstract}

\maketitle

\section{Introduction}

Many post Hartree-Fock(HF) methods utilize a localized description of the HF
wavefunction in order to increase computational
efficiency.\cite{hampel1996,saebo/1987} This is often done by a posteriori
localization methods where a unitary matrix is found that transforms the HF
orbitals into a more local representation. Some well known methods are the
Boys' localization\cite{boys_1960} and the Pipek-Mezey\\
localization\cite{pipek_1989}, just to name a couple. Despite their common use,
these localization schemes suffer from two deficiencies. First is that the HF
orbitals need to be obtained, which is undesirable for larger systems. Second,
while the orbitals produced are largely localized, they will contain tails
delocalized across the molecule. These tails can of course be removed, but
there is no guarantee that the effect on the quality of the wavefunction will
be insignificant. Instead, we propose the use of a priori localization schemes
to obtain wavefunctions suitable for post-HF methods. These localization
schemes involve subdividing the molecule into subspaces of atoms, called
fragments, and restricting orbitals to basis functions centered in those
fragments. Each fragment can contain multiple orbitals that do not need to be
orthogonal to each other. The variational parameters of the wavefunction are
then optimized with these constraints. We believe that we can obtain a more
compact form of the wavefunction that is well suited for post HF methods using
this localization scheme.

This localization method was first proposed by Adams and Gilbert\cite{adams1961, gilbert1974}
and is now referred to by a variety of names including extremely localized
molecular
orbitals(ELMO)\cite{fornili2003,genoni2008,meyer2016,meyer2016a,sironi2007,couty1997},
non-orthogonal localized molecular orbitals(NOLMO), and absolutely localized
molecular orbitals(ALMO)\cite{khaliullin2006,khaliullin2007}. Here, we will
refer to them as Strictly Localized Molecular Orbitals(sLMO). Much of the work
on these localized orbital methods in recent years has been focused on making HF
calculations on large systems more efficient with an emphasis on
proteins\cite{meyer2016,macetti2019}. In particular, this method has been used
to take advantage of both the transferability of the wavefunction as well as the
inherent computational cost reduction that sLMO-HF provides. Transferability
refers to the ability to take orbitals obtained for a molecule and use the same
orbital coefficients for another orbital in a different molecule, effectively
"transferring" the orbital from one molecule to another. Transferability is not
useful in standard HF because the canonical HF orbitals are delocalized across
the entire molecule, so the orbitals may not be reliably used to describe
orbitals in other molecules, even molecules with similar structure. Whereas
with sLMO-HF, since each orbital spans a small subset of atoms, the orbital can
be transferred to another molecule with the same atoms in a similar chemical
environment. In the context of proteins, this means defining fragments for each
amino acid, then transferring those fragmented amino acid orbitals onto the
larger protein, forming a block diagonalized coefficient matrix. The
computational cost reduction comes from the localization of the orbitals. If two
basis functions are not used together in any of the fragments, then those basis
functions will be orthogonal by definition. Therefore, terms in the one and two
electron integrals that involve the overlap between them can be ignored. This
can substantially reduce the effort required to build the Fock matrix and
thereby speed up the calculation. The degree that the calculation speeds up is
directly related to the choice of fragmentation, where more compact
fragmentation patterns result in larger cost reductions when building the Fock
matrix.

This compact choice of fragmentation will also result in a reduction of the
basis set superposition error(BSSE). Since each fragment is strictly limited in
the basis functions that it can use, BSSE between fragments can be explicitly
removed. This has been shown in the literature to be an effective way to deal
with both intermolecular\cite{khaliullin2007} and
intramolecular\cite{fornili2003} BSSE. Of course, the more compact the
fragmentation, the less BSSE that will be present in the calculation. A
reduction of BSSE also means that sLMO-HF is less sensitive to the choice of
basis set, which will be explored later in this paper.


In this paper, we present a quasi-Newton optimization method for sLMO-based
calculations. We then show that the convergence of sLMO-HF energies to the
basis set limit is exponential with respect to the $\zeta$ of the basis set
used, just as in standard HF. Yet, keeping the fragment definition constant the
sLMO-HF and the HF complete basis set(CBS) limits do not coincide. The third
purpose of this work is to show why this is the case: sLMO-HF orbitals lack
orbital penetration into neighboring nuclei, which HF orbitals do possess. It
should be pointed out that these delocalization are not artificial
'orthogonalization tails' but rather are critical in order to obtain an optimal
description of the orbital shape.  In order to develop such orbital
penetration, basis functions centered around the neighboring atoms are needed.
This does not mean that sLMO-HF calculations are doomed, but rather that
limiting the size of the orbital basis for a given orbital involves a
truncation of physical effects. One then has to choose the fragment size
according to the error that is acceptable. Lastly, we discuss the effects of
different choices of fragmentation and basis set, emphasizing how these affect
the quality of the wavefunction as well as the efficiency of the method.

\section{Theory}
Here, we will use indices $i,j,k,l$ for occupied molecular orbitals(MO),
indices $a,b,c,d$ for virtual molecular orbitals and $p,q,r,s$ for atomic
orbitals(AO).  The main idea of sLMO-HF is to optimize the orbitals such that
each MO uses some subset of the AO, written in terms of fermionic operators we
have:
\begin{equation}
  a^\dagger_i = \mathcal{N}_i\sum_p^MC_{pi}c_p^\dagger, p\notin[i]_{occ} \Rightarrow C_{pi}=0
\end{equation}
where $a^\dagger_i$ and $c_p^\dagger$ are the creation operators for the MO and AO
respectively, $C$ are the MO coefficients, $[i]_{occ}$ is the subset of the AOs
to be used for orbital i, M is the total number of basis functions and
$\mathcal{N}_i$ is a normalization constant. The definition of $[i]_{occ}$ is
arbitrary in principle, but should be chosen with chemical intuition in mind
which we will elaborate about below. We will often refer to the choice of
$[i]_{occ}$ as the fragmentation of the molecule. 

The sLMO-HF wavefunction is then simply:
\begin{equation}
  |0\rangle = \left(\prod_ia_i^\dagger\right)|-\rangle
\end{equation}

To optimize these orbitals, we choose to use a Thouless rotation. Here we
outline the approach, for details we direct the reader to the literature.
\cite{carlos} In the general case, a Thouless rotation updates the orbitals
from the initial guess, $|\Phi_0\rangle$, as
\begin{equation}
  |0\rangle = e^Z|\Phi_0\rangle
\end{equation}
with
\begin{equation}
  Z = \sum_{ia}Z_{ia}a_a^\dagger a_i
\end{equation}
The values of $Z_{ia}$ are chosen to minimize the energy:
\begin{equation}
  E = \frac{\langle\Phi_0|e^{Z^\dagger}He^Z|\Phi_0\rangle}
  {\langle\Phi_0|e^{Z^\dagger}e^Z|\Phi_0\rangle}
\end{equation}

The application of the Thouless rotation transforms the MO coefficients through
the intermediate:
\begin{align}
  \tilde{C}_i &= C_i^0 + \sum_a Z_{ia}C_a^0\\
  \tilde{C}_a &= C_a^0 - \sum_i Z_{ia}C_i^0
\end{align}
where $C_i$ and $C_a$ are the $i$th and $a$th column of the coefficient matrix. The
final orbitals are then found through the transformation:
\begin{align}
  (C_{occ})_{pi} &= \sum_j(\tilde{C}_{occ})_{pj}(L^{-1})_{ji}\\
  (C_{vir})_{pa} &= \sum_b(\tilde{C}_{vir})_{pb}(M^{-1})_{ba}
\end{align}
where $L$ and $M$ are found through the Cholesky decompositions:
\begin{align}
  1 + Z^TZ &= LL^\dagger\\
  1 + ZZ^T &= MM^\dagger
\end{align}

The above set of equations make the assumption that all of the occupied
orbitals are being rotated against the same set of virtual orbitals. This will
not work for our purposes since the domain of basis functions of each occupied
orbital is different. To circumvent this problem, we can construct a set of
virtual orbitals for each occupied orbitals, which can be done as follows. Let
$A$ be a set of orthogonal orbitals that span the set of AO $[i]_{occ}$ for
some occupied orbital $C_i$. We can find a basis that contains $C_i$ as one of
the basis vectors using the singular value decomposition:
\begin{equation}
  A^\dagger SC_i = U\Sigma V^\dagger
\end{equation}
where $S$ is the AO overlap matrix. We then define the matrix:
\begin{equation}
  B = AU
\end{equation}
The matrix $B$ will have the occupied orbital in the first column and the
remaining vectors will be a set of orthogonal vectors that span $[i]$ and will
act as a set of virtual orbitals to rotate $C_i$ against. The above procedure
for applying the Thouless rotation can be applied to each orbital where an $L$
and $M$ matrix can be stored for each which we will denote as $L_i$ and $M_i$.

Once we have the updated orbitals, we can calculate the density as 
\begin{gather}
  \tilde{S} = C_{occ}^\dagger S C_{occ}\\
  \rho = C_{occ} \tilde{S}^{-1} C_{occ}^\dagger
\end{gather}
The Fock matrix, $F$, and energy are then evaluated in the standard way.

The last thing we need to calculate is the gradient. Since we are using a
quasi-Newton optimization, we will first calculate the local gradient on the
tangent space of the current step, then transform the local gradient back to
the tangent space of the initial point. The local gradient has been derived by
Stoll et al.\cite{stoll_1980} For some orbital $C_i$, the local gradient is
given by:
\begin{equation}
  g_i = 4C_{vir}^\dagger(1-S\rho)FC_i\tilde{S}^{-1}_i
\end{equation}
where $\tilde{S}_i^{-1}$ is the $i$th column of the inverse overlap matrix.
The global gradient is then calculated by
\begin{equation}
  G_i = (M_i^\dagger)^{-1} g_i(L_i)^{-1}
\end{equation}

\section{Scaling}
Since we intend to use these orbitals for local correlation techniques, we need
to ensure that the method scales linearly for large systems. The linear scaling
of the method stems from the observation that the density of each orbital
decays exponentially away from the basis functions in $[i]$. This is guaranteed
by the fact that we place explicit zeros for coefficients on AO outside $[i]$.
In particular, if we have the density written as:
\begin{equation}
  \rho_{qs} = (C_{occ})_{qi} (\tilde{S}^{-1})_{ij} (C_{occ}^\dagger)_{js}
\end{equation}
we have that $q\in [i]_{occ}$ and $s\in [j]_{occ}$. Further, the inverse
overlap matrix requires that orbital $i$ is close to orbital $j$. These
together tell us that the density matrix is banded with band width determined
by the fragmentation choice. With this the most expensive part of the calculation is
the construction of the Fock operator, which includes the Coulomb contraction:
\begin{equation}
  \sum_{qs}\langle pq|rs\rangle \rho_{qs}
\end{equation}
Here, we have that $p$ and $r$ must be close due to the overlap in the
integral. Lastly, we have that the $\frac{1}{r}$ operator requires that
orbitals $p$ and $r$ must be close to $q$ and $s$. Taken together, the Fock
matrix then becomes banded as well, producing the linear behavior for large
systems.

\section{Implementation}
We performed calculations using in-house code using Gaussian 16\cite{g16} to
evaluate the Fock matrix and overlap integrals. We used the L-BFGS algorithm to
optimize the orbitals with a convergence of $10^{-5}$ in the norm of the
gradient. The initial guess for the orbitals was taken from an RHF calculation
whose orbitals were localized with the Boys function and basis functions
outside the fragmentation projected out.  We often found it useful to optimize
the sLMO-HF wavefunction in a minimal basis, then project the orbitals into a
larger basis to produce a better initial guess.

As we will discuss below, convergence of the sLMO-HF orbitals can be difficult
due to the non-orthogonality of the orbitals and more specifically that the
virtual spaces for each occupied orbital is not orthogonal to the other
occupied orbitals, as doing so would require that the orbitals is expanded
outside of the fragmentation. There are instances where projecting out other
occupied orbitals can be done, specifically if $[j]\subset [i]$, then $C_j$ can
be projected out of the virtual space for $C_i$.

\section{Fragmentation}
The fragmentation for a molecule is in principle arbitrary, however if chosen
poorly will likely lead to a very poor result if it converges at all. We will
discuss several fragmentations below, but the all follow a similar motif. When
the canonical HF orbitals are localized, they produce an orbitals for each
bond, lone pair and core electron. The lone pair and core electron orbitals
tend to be primarily localized on one atom and the bond orbitals tend to be
primarily localized on two neighboring atoms.

\begin{figure*}
  \begin{center}
    \includegraphics*[width=\textwidth]{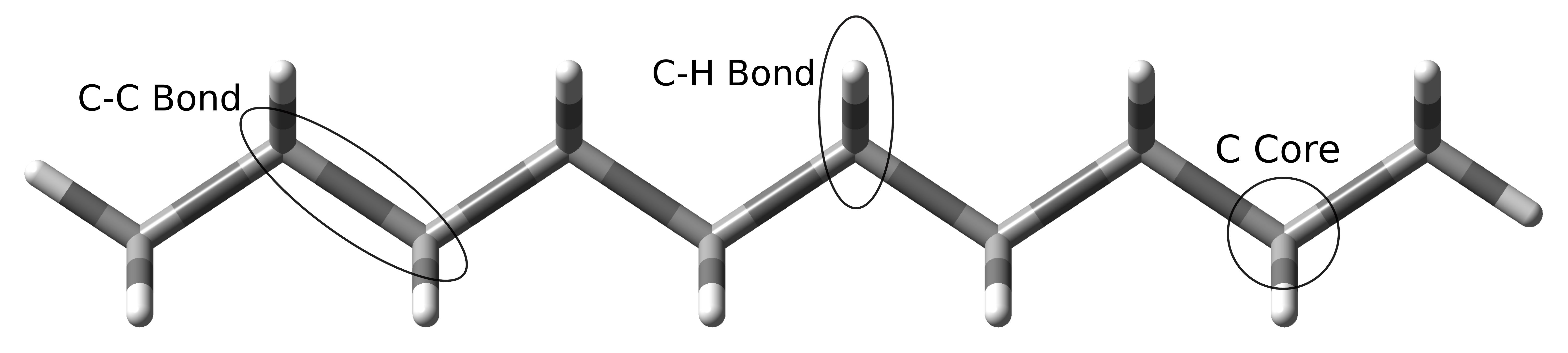}\\
    \includegraphics*[width=\textwidth]{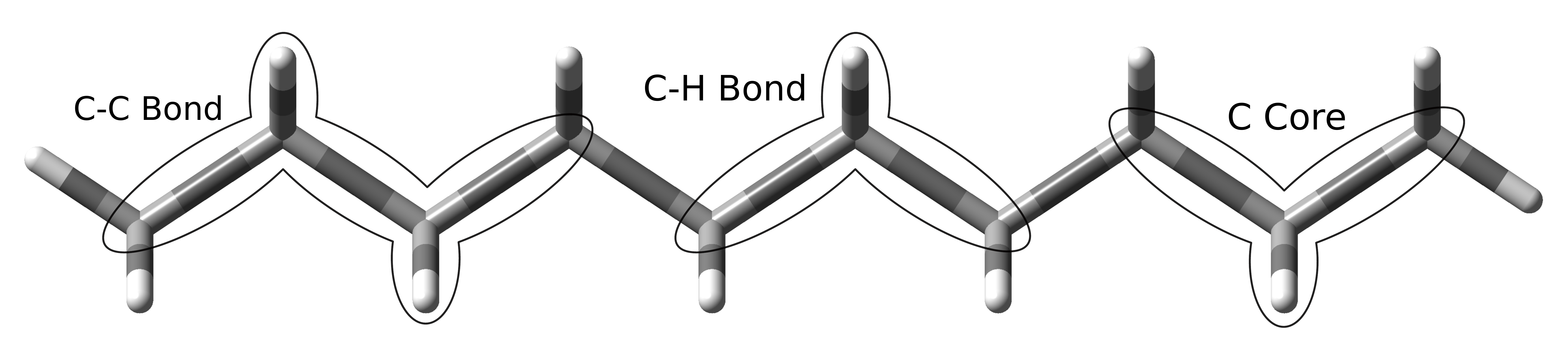}
  \end{center}
  \caption{Minimal(top) and extended(bottom) fragmentations shown for the
  decane molecule.}
  \label{fig:dec_frag}
\end{figure*}

With this knowledge, we can define a minimal fragmentation to choose the
fragmentation of a lone pair and core electron to use the basis functions from
that one atom and the fragmentation for a bond to use basis functions on the
two atoms of the bond. We can produce an extended fragmentation by expanding
each fragment to the atoms that are bonded to the atoms in the minimal
fragmentation. A visual representation of these fragmentations are shown in
Fig. \ref{fig:dec_frag}. Naturally, these are not the only choices but they are
the simplest. We will discuss the effects of these choices below.

Alternatively, we can use more automated approach as follows. First perform a
HF calculation on the molecule using a minimal basis and localize the orbitals
in the usual way. Then for each orbital, included the basis functions on any
atom whose Mulliken weight is above some value. We found that a value of 0.4
works well for all of the systems that we studied.

\section{Results for sLMO-HF}
{\em decane geometry}. We use a generic geometry for decane determined by only a
few geometrical parameters. All C-C (C-H) bond lengths are set to $1.53$
($1.09$)~\AA, and all C-C-C angles are set to $113$~deg. Each C has a pair of H
atoms symmetrically placed on a plane normal to the C backbone plane, with an
angle H-C-H of $107$~deg. The two remaining H atoms (one on each of the terminal
C's) are placed on the same plane as the C backbone, with angles H-C-C set to
$113$~deg.

\subsection{Complete Basis Set Limit}
\begin{figure}
  \centering
  \includegraphics[width=\columnwidth]{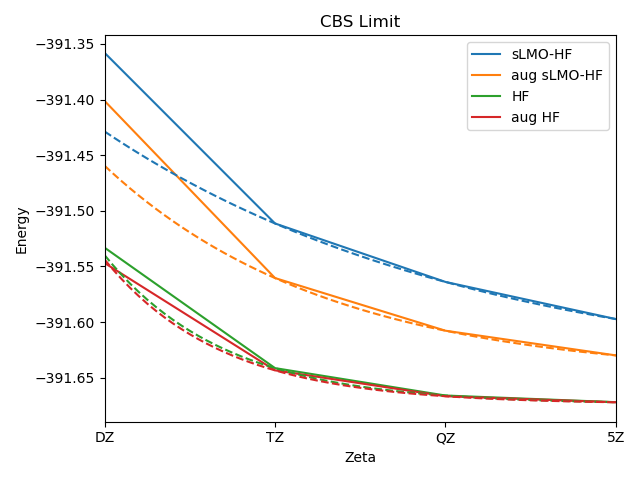}
  \caption{Energies (in a.u.) for sLMO-HF and HF for decane using cc-pVNZ basis
  sets. Dotted lines are the extrapolated curves.
  \label{fig:decane_CBS}}
\end{figure}

A calculation that is desirable and well facilitated by the use of sLMO-HF is
the determination of the energy at the CBS limit. Here, we show that the
sLMO-HF energy converges exponentially with basis set $\zeta$ in the same way
as HF using decane. The fragmentation we choose is the minimal one described
above where we restrict core electrons to basis functions centered on the
corresponding carbon and restrict the valance electrons to basis functions
centered at adjacent atoms for each bond. In Table \ref{tab:decane_CBS}, we
report the results of an exponential extrapolation of sLMO-HF and HF to the CBS
limit using TZ, QZ, and 5Z basis sets. Figure \ref{fig:decane_CBS} shows the
sLMO-HF and HF CBS energies as a function of $\zeta$. Here we see that both
sLMO-HF and HF converge exponentially, but HF has a lower CBS limit than
sLMO-HF. This is to be expected since the quality of the sLMO-HF wavefunction
will be lower than the HF wavefunction due to having fewer variational
parameters and no basis functions outside of the fragment which results in the
sLMO-HF wavefunction losing several features that are critical for an accurate
description of the electronic structure. This point will be discussed further
in the next section. Though in principle, if a complete basis set was used for
each fragment, the sLMO-HF wavefunction should be identical to the HF
wavefunction, therefore the energy should be the same.

\begin{table}
  \centering
  \caption{HF energies (in a.u.) calculated for the decane molecule with
    various basis sets
    set. \label{tab:decane_CBS}}
  \footnotesize
  \begin{tabular}{@{}l r r r r@{}}
  Basis Set & sLMO-HF & aug sLMO-HF & HF & aug HF\\
  \hline
  cc-pVDZ  & -391.358067 &  -391.401287 &  -391.533203 & -391.547006 \\
  cc-pVTZ  & -391.511417 &  -391.560495 &  -391.641450 & -391.643464 \\
  cc-pVQZ  & -391.563959 &  -391.607818 &  -391.666082 & -391.666718 \\
  cc-pV5Z  & -391.597352 &  -391.630004 &  -391.672060 & -391.672149 \\
  \hline
  CBS      & -391.655584 &  -391.649585 &  -391.673975 & -391.673803 \\
  \end{tabular}
\end{table}

\subsection{Comparison with Canonical HF}

For the following discussion, we denote by $|\Phi \rangle$ the (converged)
sLMO-HF wavefunction for decane, and by $|\Psi \rangle$ the corresponding HF
wavefunction. We build two alternate representations of $|\Psi \rangle$ as
detailed below.

\begin{itemize}
  \item For each orbital $\phi_i (\mathbf{r})$ in $|\Phi \rangle$, we construct
    a corresponding orbital $\tilde{\psi}_i (\mathbf{r})$ which resembles
    $\phi_i$ the most but is part of the occupied HF space in $|\Psi \rangle$.
    This is done by choosing $\tilde{\psi}_i$ as the single eigenvector with
    non-zero eigenvalue of the matrix $M_{\alpha \beta} = \langle \alpha |
    \phi_i \rangle \, \langle \phi_i | \beta \rangle$, where $\{\alpha \}$ span
    the HF occupied orbital space. We finally let $|\Psi \rangle_\mathrm{noLMO}
    = \mathcal{A} \, \tilde{\psi}_1 \tilde{\psi}_2 \cdots \tilde{\psi}_N$, where
    ``noLMO'' stands for non-orthogonal Localized MOs, given that the $\psi$
    orbitals are non-orthogonal amongst each other, they are localized by
    construction, and they fully span the HF wavefunction, i.e., $|\Psi \rangle$
    and $|\Psi \rangle_\mathrm{noLMO}$ are the same wavefunction up to a
    normalization and phase factors.

  \item We then carry a L\"owdin symmetric orthogonalization of the orbitals
    $\{ \tilde{\psi} \}$ to construct a new set of orbitals $\{ \bar{\psi} \}$
    which are orthonormal yet remain localized (by virtue of the symmetric
    orthogonalization). We denote $|\Psi \rangle_\mathrm{LMO} = \mathcal{A} \,
    \bar{\psi}_1 \bar{\psi}_2 \cdots \bar{\psi}_N$ where ``LMO'' stands for
    (orthonormal) Localized MOs. It is again true that $|\Psi \rangle$ and
    $|\Psi \rangle_\mathrm{LMO}$ are the same wavefunction up to an arbitrary
    phase factor.
\end{itemize}

Table \ref{tab:decane_energy} below shows the energy (and its splitting by
contribution) of $|\Phi \rangle$, $|\Psi \rangle$, $|\Psi
\rangle_\mathrm{LMO}^\mathrm{td}$, and $|\Psi
\rangle_\mathrm{noLMO}^\mathrm{td}$. The last two wavefunctions result from
``tail deletions''\footnote{Here, ``tail deletion'' can be done by zeroing out
the appropriate coefficients in each orbital. We use a somewhat more elegant
approach where each orbital is projected into the basis functions $\{\chi\}$
that would be used for such orbital.}  out of $|\Psi \rangle_\mathrm{LMO}$, and
$|\Psi \rangle_\mathrm{noLMO}$, respectively. Deleting the tails out of LMOs
increases the energy, with respect to the HF one, by $\approx 410$~mhartree, or
$\approx 6.7$~mhartree per valence electron. On the other hand, tail deletion
out of noLMOs only increases the energy by $\approx 3.1$~mhartree per valence
electron. Relaxing $|\Psi \rangle_\mathrm{noLMO}^\mathrm{td}$ down to $|\Phi
\rangle$, where the wavefunction is optimized in the presence of the chosen
constraints, only results in a $15.3$~mhartree (or $0.25$~mhartree per valence
electron) further lowering in the energy. While it has been pointed out before
\cite{sironi2007} that $|\Phi \rangle$ is significantly below in energy than $|\Psi
\rangle_\mathrm{LMO}^\mathrm{td}$ suggesting the importance of the relaxation of
the wavefunction in the presence of the constraint, we see that the energy gain
is relatively small if one works with a non-orthogonal localized representation
of the HF wavefunction from the onset.

\begin{table}
  \centering
  \caption{Energies (in a.u.) calculated for the decane molecule with
    various wavefunctions using the cc-pVDZ basis
    set. \label{tab:decane_energy}}
  \footnotesize
  \setlength{\tabcolsep}{1pt}
  \begin{tabular}{@{}l r r r r r@{}}
  wavefunction & total & kinetic & nuc-elec & coulomb & exchange \\
  \hline
  HF           & -391.5332 &  391.2686 &  -1947.2560 &  705.5237 &  -59.8943 \\
  sLMO-HF      & -391.3581 &  391.7844 &  -1947.8067 &  705.7785 &  -59.9391 \\
  HF(LMO)/td   & -391.1201 &  394.4045 &  -1952.4958 &  708.5260 &  -60.3797 \\
  HF(noLMO)/td & -391.3428 &  392.3262 &  -1949.5289 &  707.0339 &  -59.9989 \\
  \end{tabular}
\end{table}

\begin{figure*}
  \centering
  \includegraphics[height=2.1in]{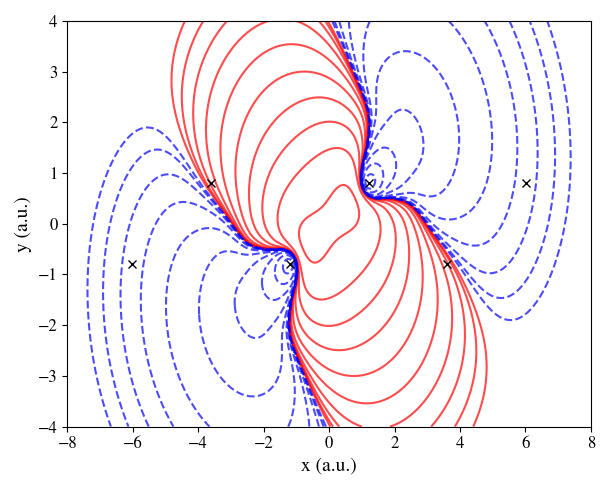}\quad
  \includegraphics[height=2.1in]{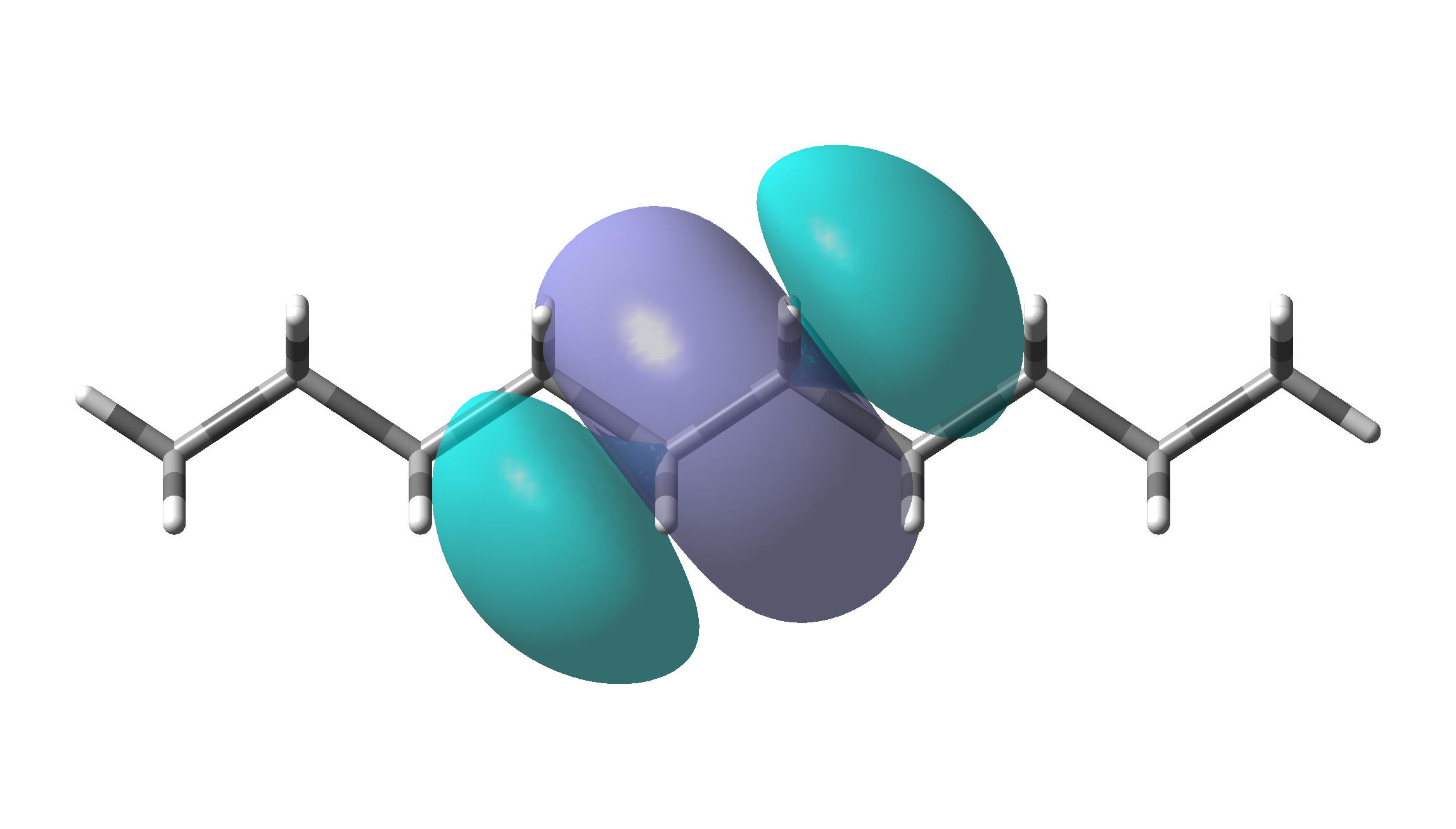}
  \\
  \includegraphics[height=2.1in]{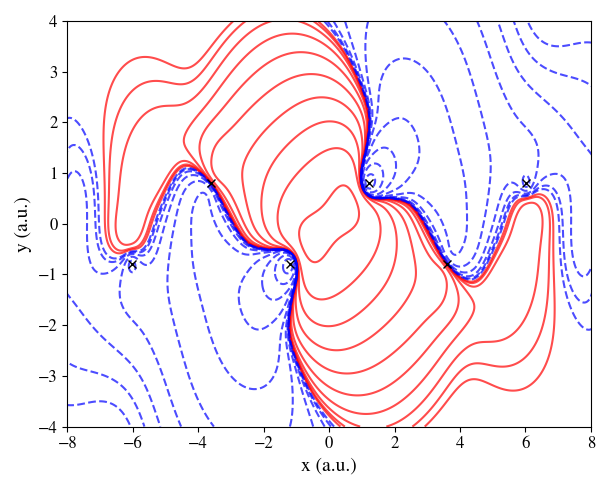}\quad
  \includegraphics[height=2.1in]{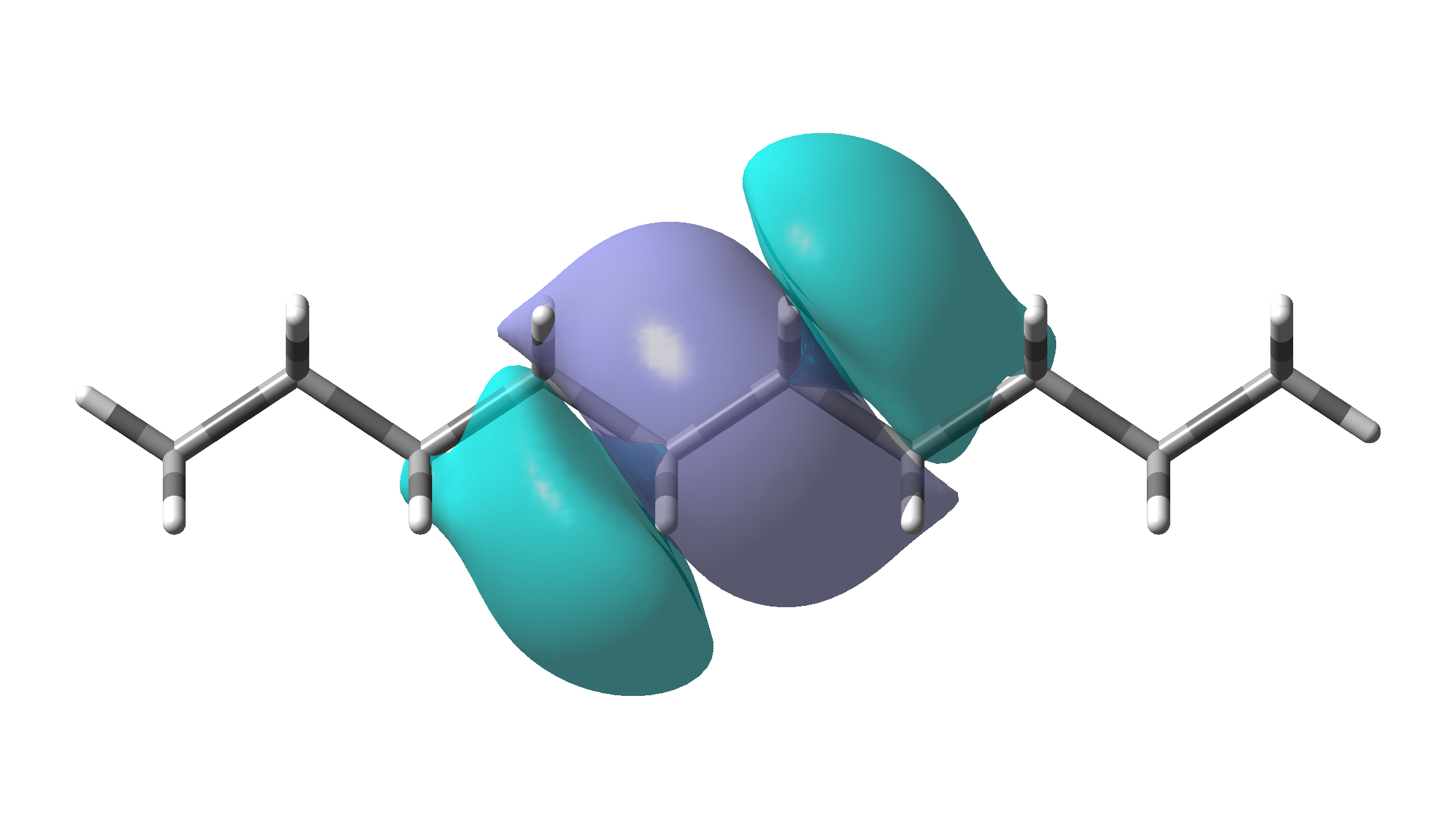}
  \\
  \includegraphics[height=2.1in]{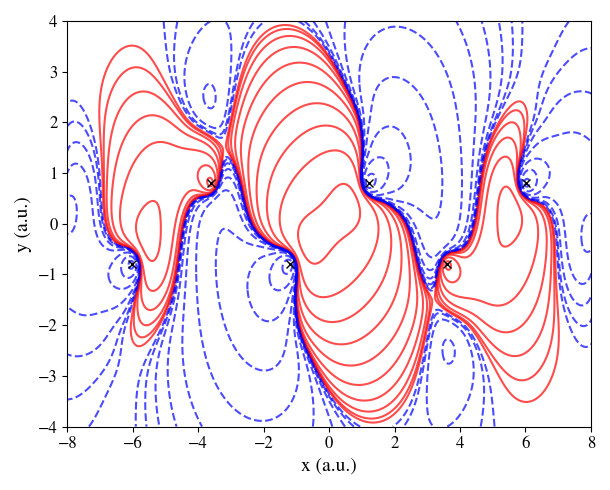}\quad
  \includegraphics[height=2.1in]{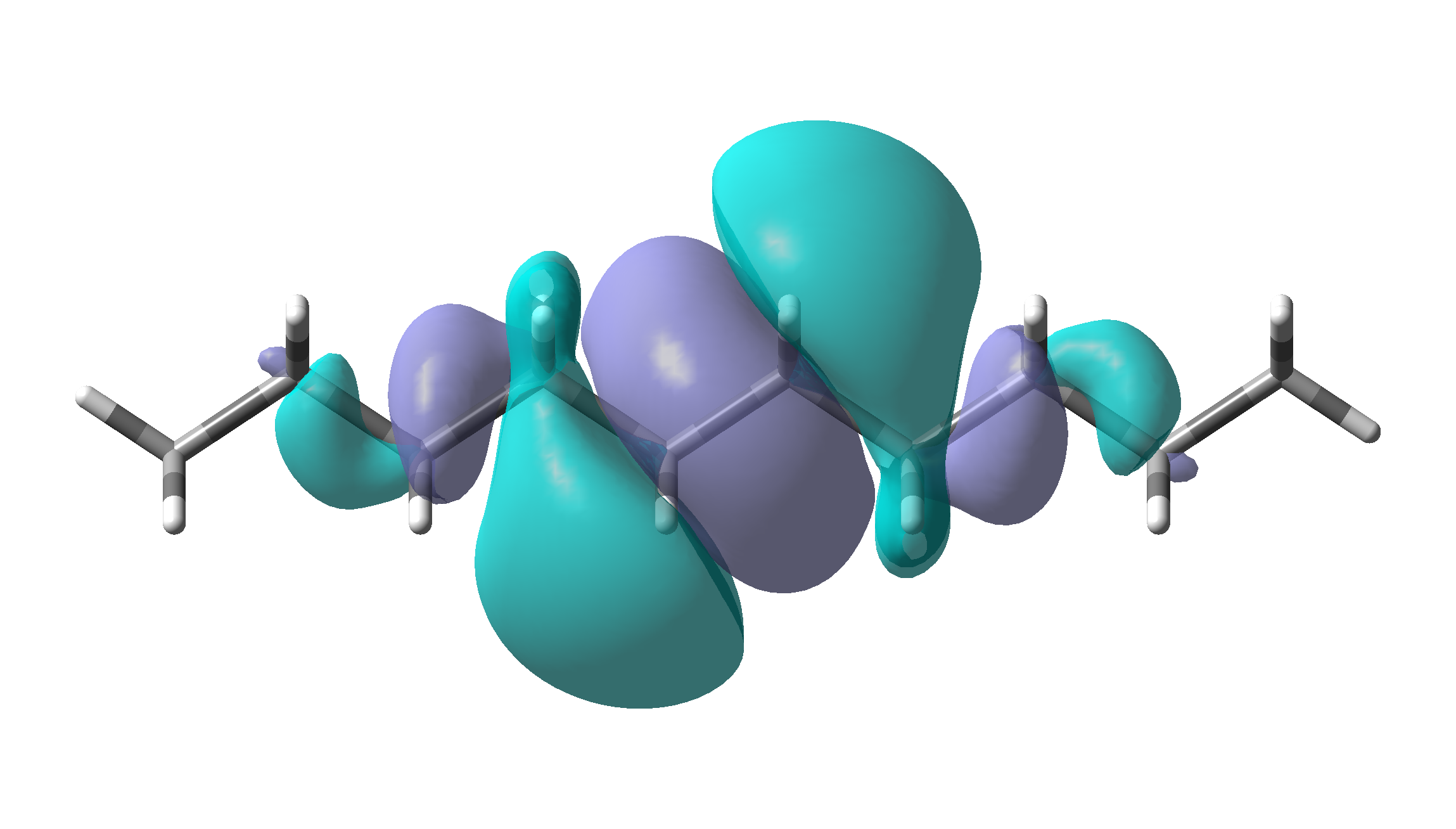}
  \caption{Contour plot and corresponding isosurfaces of the central C-C
    $\sigma$ orbital in decane computed with sLMO-HF (top) and HF (center:
    noLMO; bottom: LMO) using the cc-pVDZ basis set. The decane geometry is that
    discussed above, with the carbon backbone aligned along the x-y plane. The
    position of the carbon atoms is shown with an ``x'' mark in the contour
    plots. Contours are drawn with an equispaced logarithmic grid between
    $10^{-0.5}$ and $10^{-3.5}$~a.u. Solid and dashed lines indicate opposite
    phases. The isosurfaces were drawn at 0.003~a.u.\label{fig:decane_orb}}
\end{figure*}

We show in Fig. \ref{fig:decane_orb} contour and isosurfaces plots of the
converged C-C $\sigma$ bond orbital in $|\Phi \rangle$, $|\Psi
\rangle_\mathrm{noLMO}$, and $|\Psi \rangle_\mathrm{LMO}$. The similarity
between the sLMO and noLMO orbitals is striking: while the two are different in
regions far from the central C atoms, it is not evident that $\phi$ is more
localized than $\varphi$. A measure of their difference can be obtained by
evaluating $|\langle \phi | \varphi \rangle|$, which in this case is $0.9986$,
indicating only a small difference between the orbitals despite the observed
differences in the contour plot. The LMO, on the other hand, shows the long
tails needed to yield an orthonormal representation. In order to quantify the
real-space localization, we computed the orbital spreads defined by
\begin{equation}
  s^2_\chi = \langle \chi | {\bf r} \cdot {\bf r} | \chi \rangle -
  \langle \chi | {\bf r} | \chi \rangle \cdot \langle \chi | {\bf r} |
  \chi \rangle.
\end{equation}
We obtained $s = 1.496$ (sLMO) and $s = 1.500$ (noLMO). The
difference between the sLMO and noLMO spreads amounts to $< 0.3$~\%.

Our interpretation of the plots is as follows: both orbitals are very similar
in regions of large electron density and both have similar real-space
localization properties, but $\phi$ misses some of the fine structure of
$\varphi$ near the nuclei. For instance, the nearest neighbor C nuclei appear
as nodes in $\varphi$ but not quite in $\phi$, where the node is a bit shifted;
similarly, $\varphi$ shows a nodal feature near the second nearest neighbors
which $\phi$ completely misses. The reason for this difference is that without
basis functions centered on the neighboring atoms, it is difficult to develop
these features. Developing these features on neighboring atoms would require a
very large basis with high angular momentum functions on the atoms in the
fragment.


\begin{figure*}
  \centering
  \begin{minipage}{0.3\textwidth}
    \includegraphics[width=\textwidth]{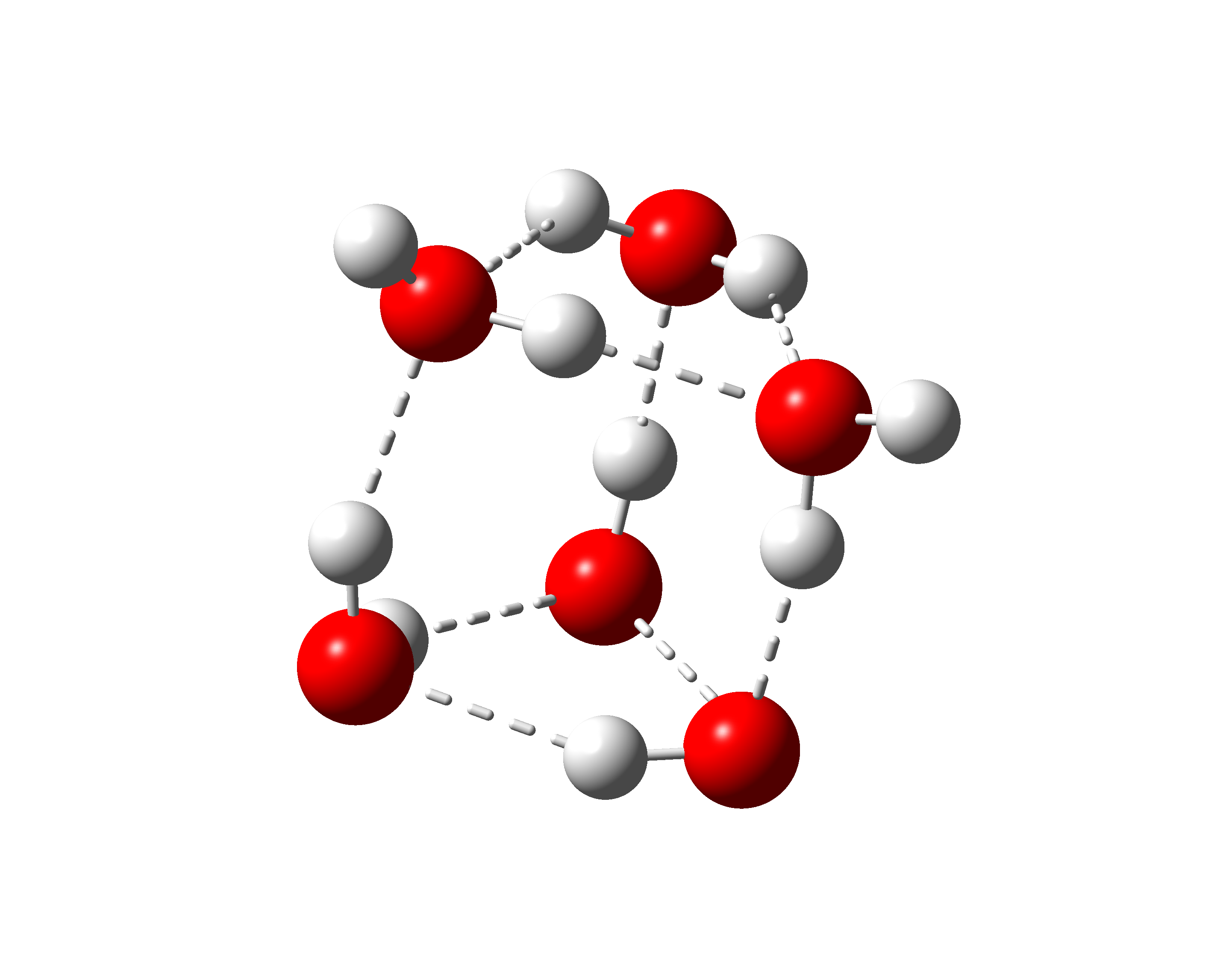}
    \center{Prism}
  \end{minipage}%
  \begin{minipage}{0.3\textwidth}
    \includegraphics[width=\textwidth]{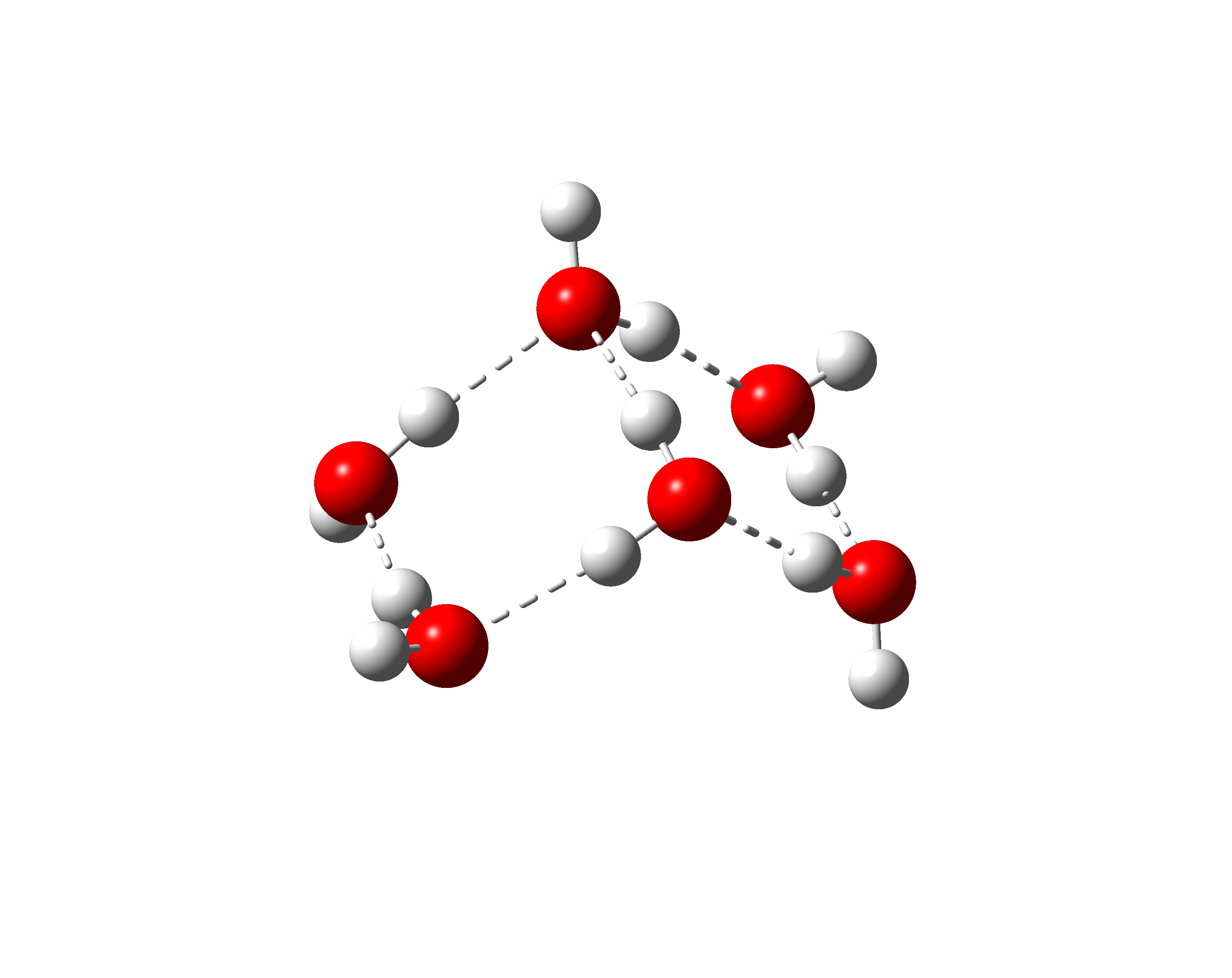}
    \center{Book}
  \end{minipage}
  \begin{minipage}{0.3\textwidth}
    \includegraphics[width=\textwidth]{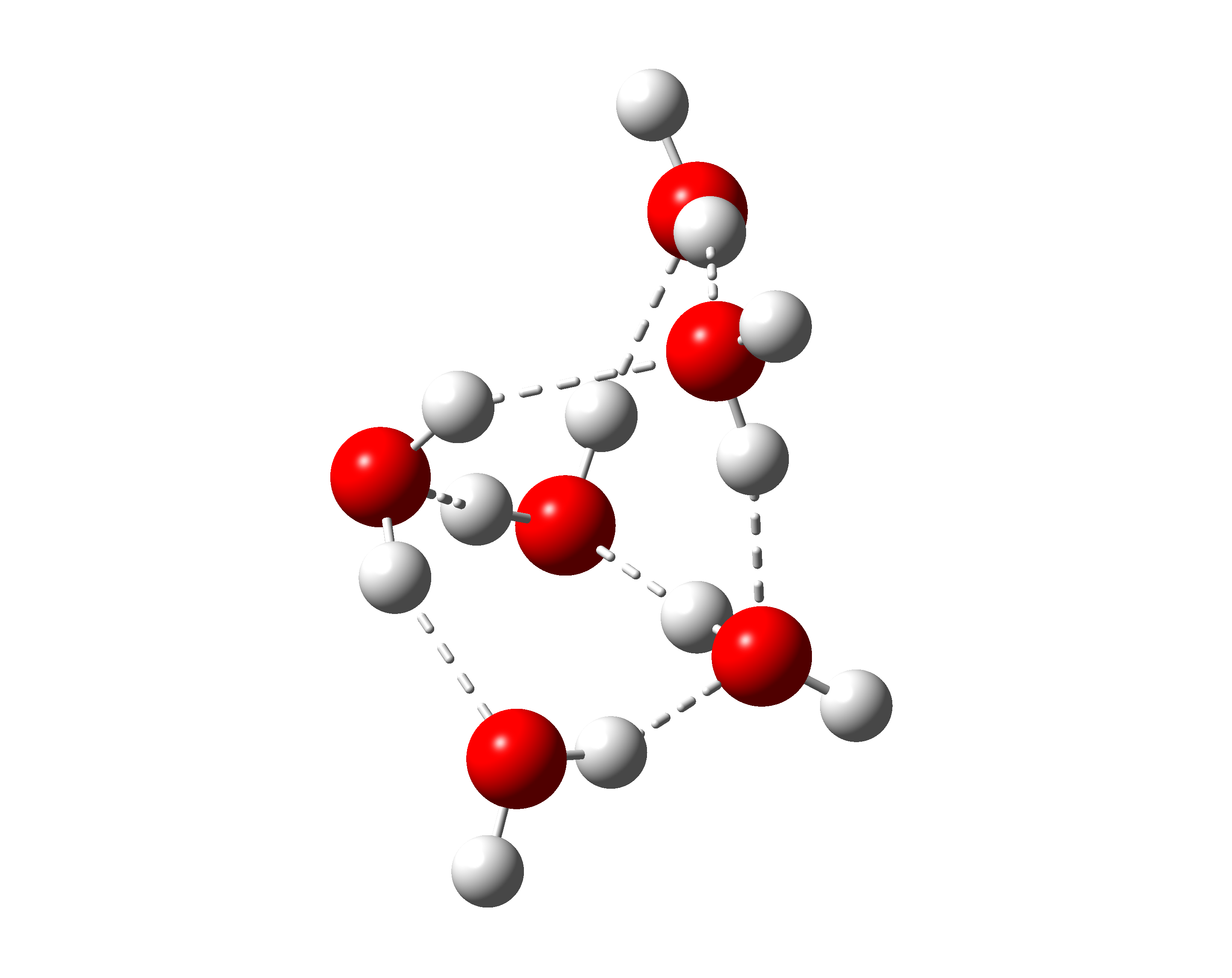}
    \center{Cage}
  \end{minipage}%
  \\
  \begin{minipage}{0.3\textwidth}
    \includegraphics[width=\textwidth]{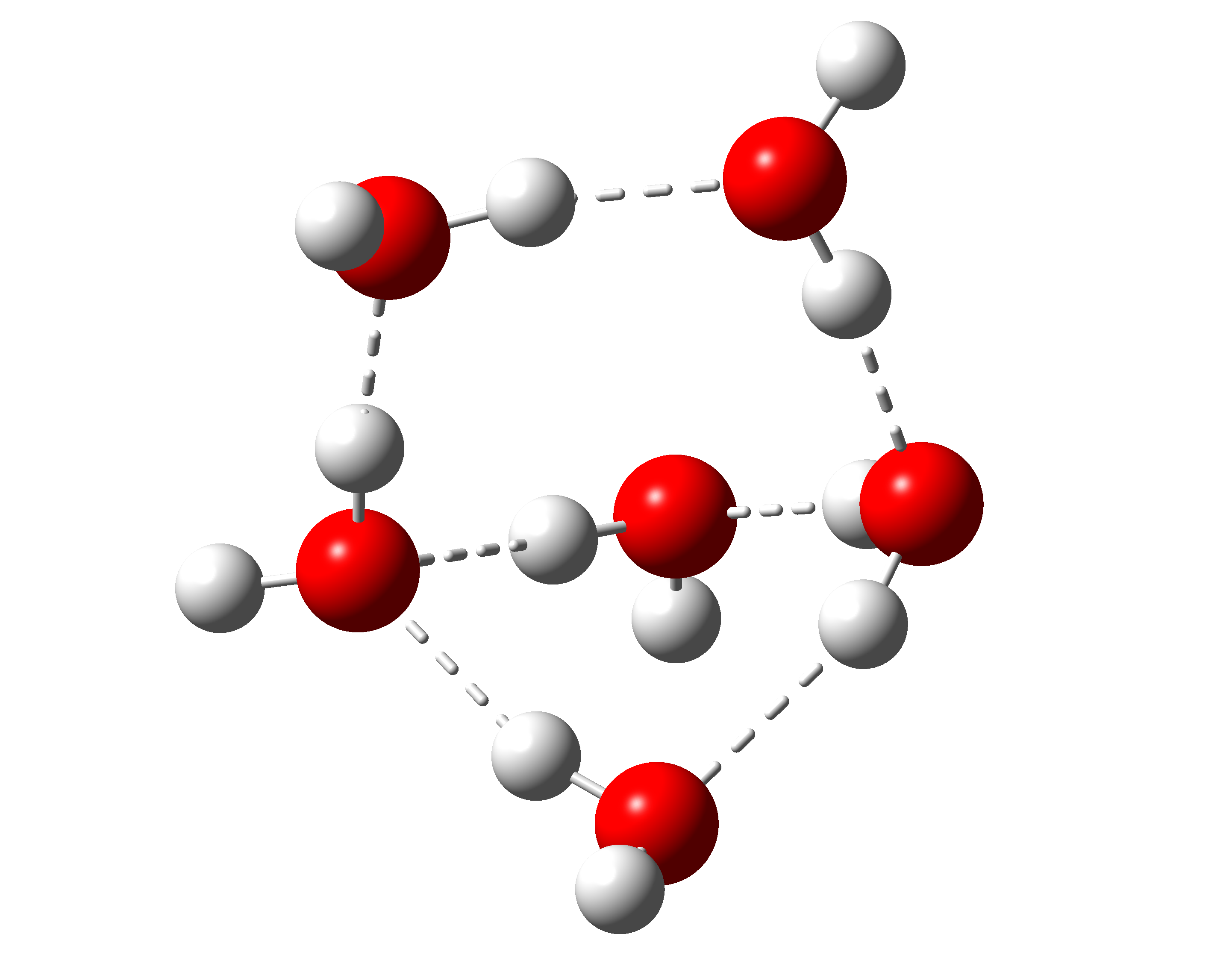}
    \center{Open Bag}
  \end{minipage}
  \begin{minipage}{0.3\textwidth}
    \includegraphics[width=\textwidth]{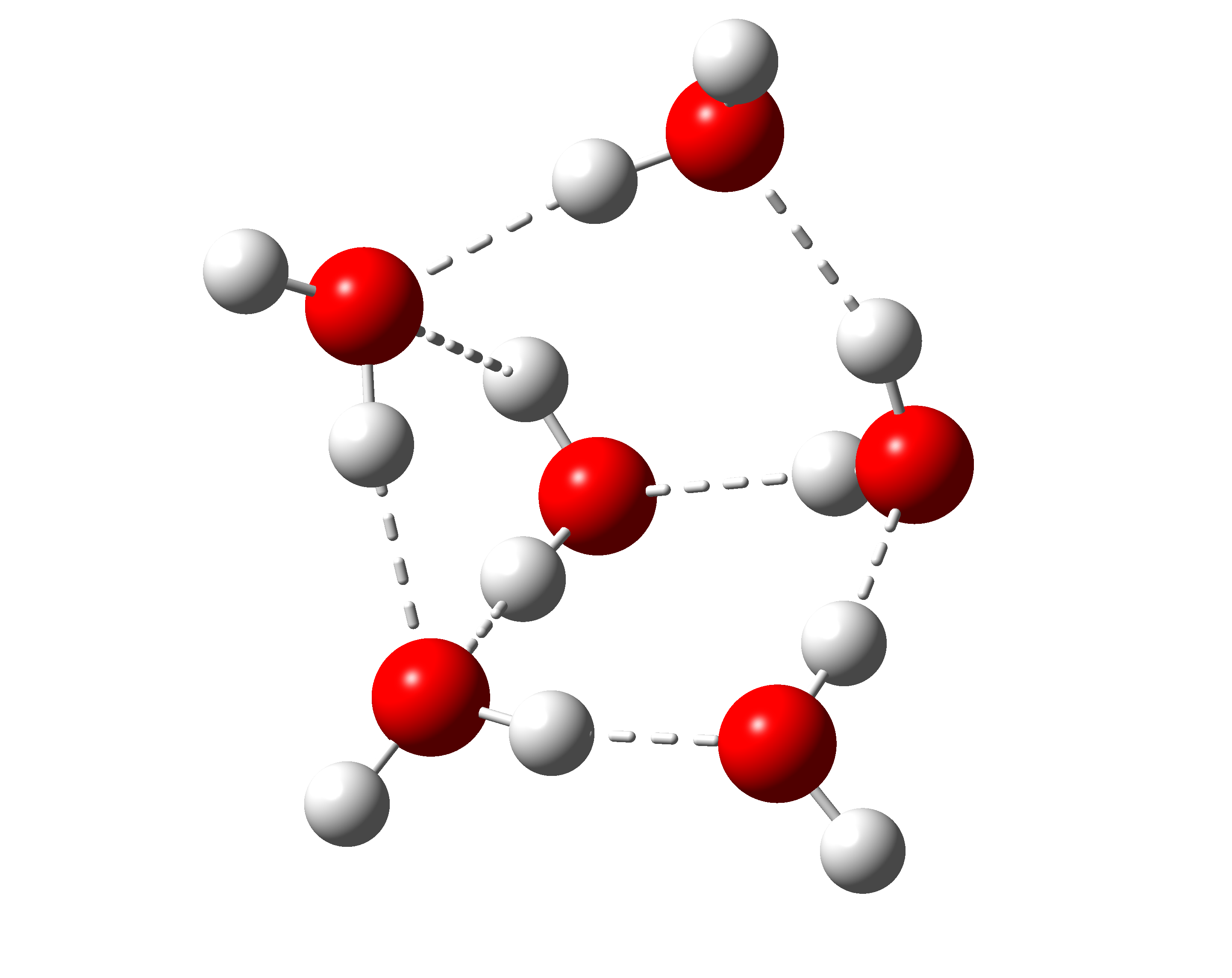}
    \center{Prism/Book}
  \end{minipage}%
  \begin{minipage}{0.3\textwidth}
    \includegraphics[width=\textwidth]{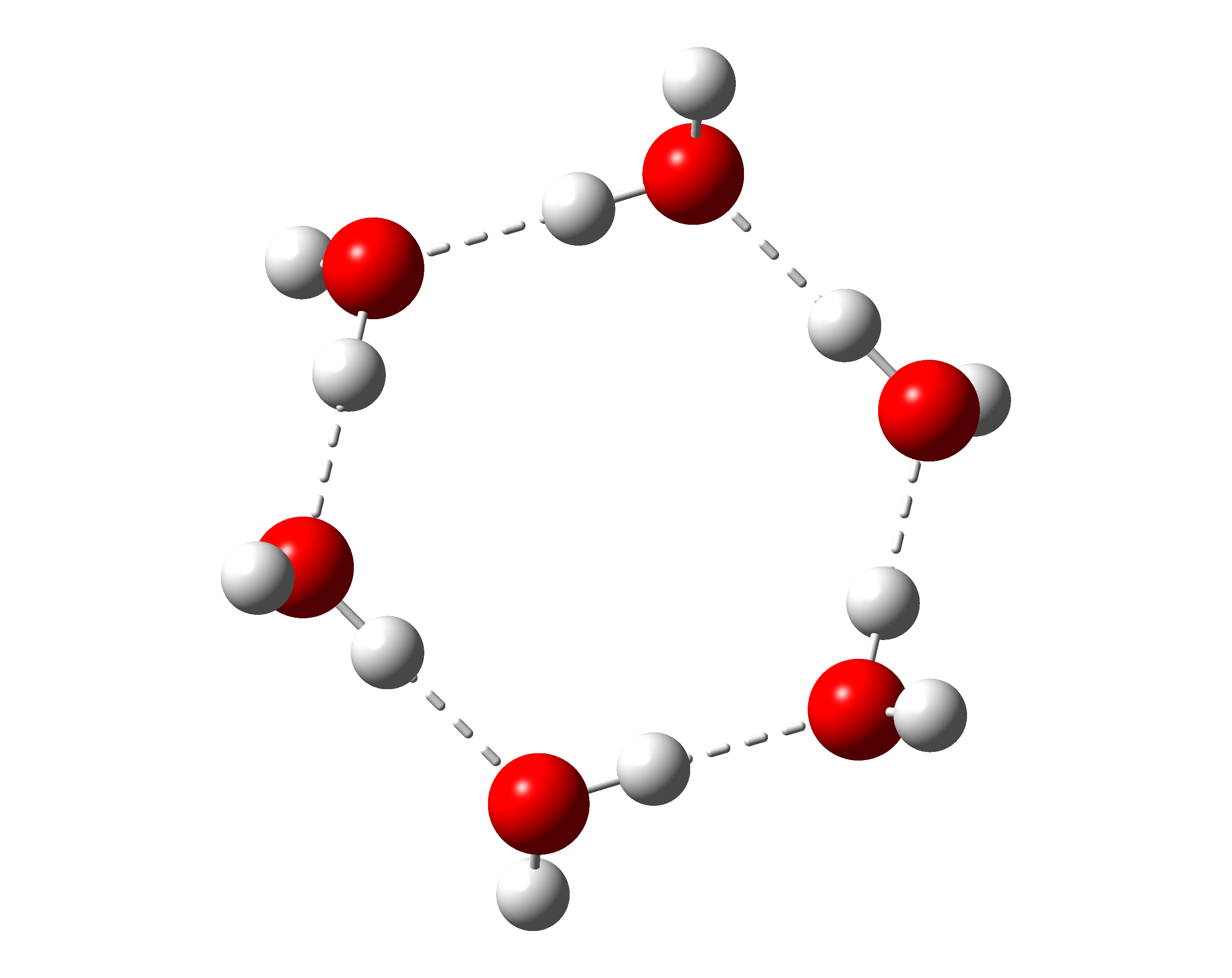}
    \center{Chair}
  \end{minipage}
  \caption{Structures for water hexamer conformations. Dotted lines indicate 
  hydrogen bonding. \label{fig:water_struct}}
\end{figure*}

\subsection{Fragmentation Choice}
In order to examine the effects of choosing different fragmentation patterns and
basis sets, it is helpful to study a more complicated system. For this, we chose
to study water hexamers in six different structural conformations with very
similar energies shown in Figure \ref{fig:water_struct}. The reason for this
choice of system is that water clusters have been extensively
studied\cite{khan2004,lenz2009,lin2005,tsurusawa1999} and are not only
physically complex, including strong intermolecular interactions and many
possible conformers, but also present a theoretical difficulty in dealing with
BSSE.  Table \ref{tab:water_energies} contains relative energies of the water
hexamers relative to the prism conformer. Here, we examine three different
fragmentation patterns. In the minimal fragmentation, we restrict the lone pair
and core electrons to basis functions centered on the corresponding oxygen. The
remaining electrons are each restricted to basis functions centered oxygen and
hydrogen atoms that are adjacent to each other for each bond. In the extended
fragmentation, we restrict both core and valance electrons to basis functions
centered within their corresponding water molecule. The hbond fragmentation
pattern is the same as the minimal fragmentation with the exception that lone
pairs and bonding electrons are extended to neighboring water molecules where
structurally appropriate for hydrogen bonding. 

\begin{figure*}
  \centering
  \begin{minipage}{0.32\textwidth}
    \center{Book}
    \includegraphics[width=\textwidth]{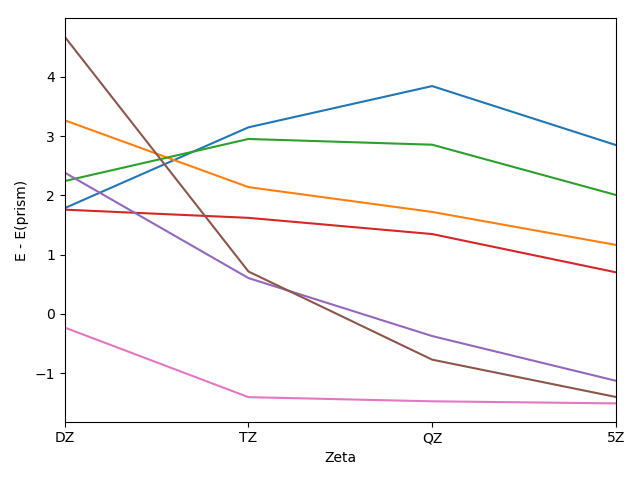}
  \end{minipage}%
  \begin{minipage}{0.32\textwidth}
    \center{Cage}
    \includegraphics[width=\textwidth]{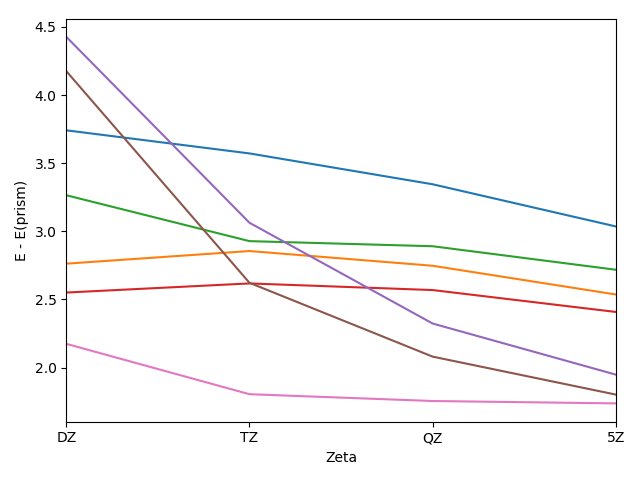}
  \end{minipage}%
  \begin{minipage}{0.32\textwidth}
    \center{Open Bag}
    \includegraphics[width=\textwidth]{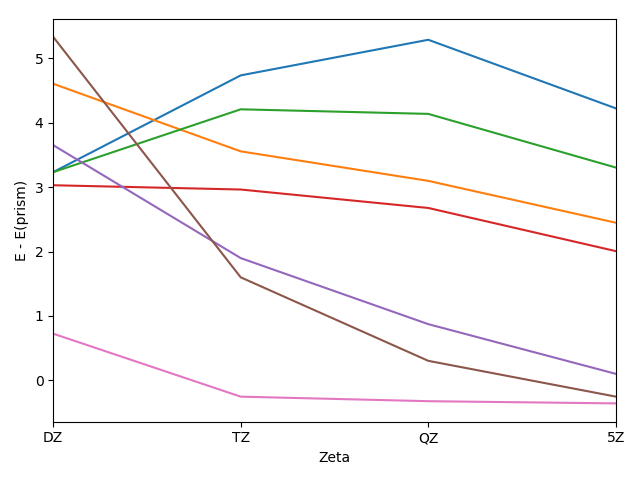}
  \end{minipage}%
  \\
  \begin{minipage}{0.32\textwidth}
    \center{Prism/Book}
    \includegraphics[width=\textwidth]{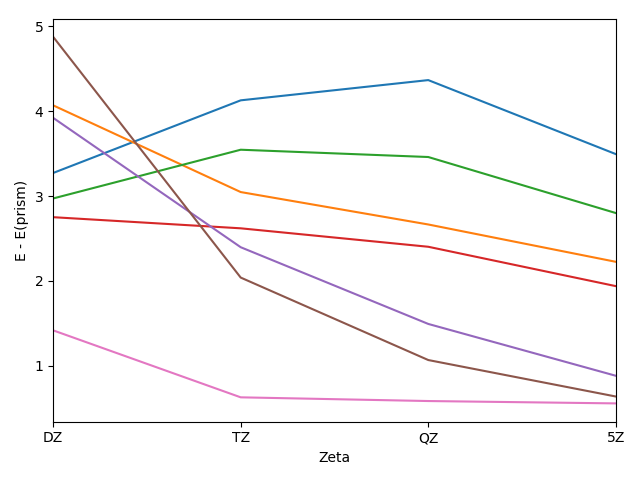}
  \end{minipage}%
  \begin{minipage}{0.32\textwidth}
    \center{Chair}
    \includegraphics[width=\textwidth]{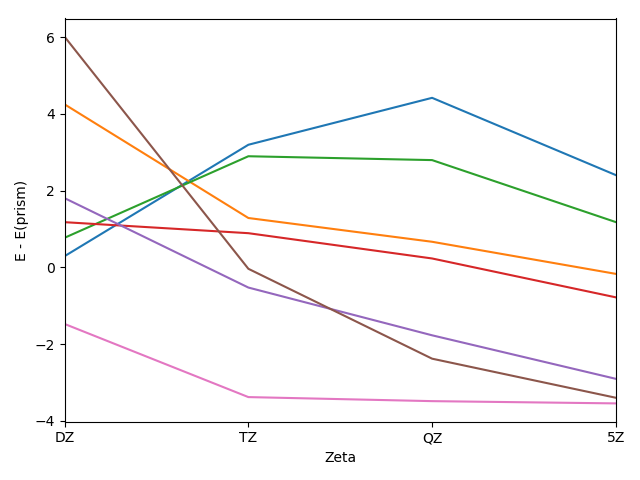}
  \end{minipage}%
  \begin{minipage}{0.32\textwidth}
    \centering
    \includegraphics[width=\textwidth]{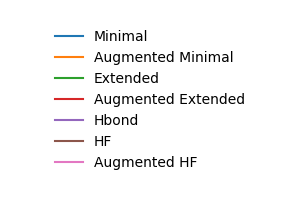}
  \end{minipage}
  \caption{Energies(mhartree) for water hexamer conformations
   relative the prism conformer. \label{fig:water_energy}}
\end{figure*}

As previously mentioned, the localization that sLMO-HF uses limits the amount of
BSSE that is present based on the fragmentation pattern used. This makes sLMO-HF
calculations less sensitive to the basis set size than HF, especially for
smaller basis sets, where the BSSE is most prominent. We can see this stability
in the relative energy of water cluster conformers shown in Figure
\ref{fig:water_energy}. Here, we see that the HF relative energies moving from
cc-pVDZ to cc-pVTZ basis sets show a change of $\approx 2-3$ mhartrees, where as
sLMO-HF energies in all but the minimal fragmentation change $<1$ mhartree. Then
when we look at the energy change from cc-pVQZ to cc-pV5Z, where BSSE is less
prominent, we see that the difference in energy for both sLMO-HF and HF are
similar. However, this stability requires an appropriate choice of fragmentation
and basis set. In the minimal fragmentation, we see larger energy changes going
from cc-pVDZ to cc-pVTZ compared to other fragmentations. We also see that the
relative energies do no change monotonically with basis set size for minimal and
extended fragmentations, noting that for several conformers, the relative energy
increases for smaller basis sets then decreases for larger basis sets. While
this issue is resolved by using an augmented basis set, these inconsistencies
indicate that these fragment choices may be too restrictive for this system and
shows that one should be careful in choosing the fragmentation that is most
appropriate for the system being studied. For water clusters, a much more
appropriate choice is the hbond fragmentation which allows the hydrogen bonding
to be accounted for. Here, we see that the relative energies for the hbond
fragmentation are $\approx 0.5$ mhartrees off of the HF energies using a cc-pV5Z
basis set where as the minimal fragmentation is $\approx 2-3$ mhartrees off for
the same basis set. So by simply extending the fragments to include hydrogen
bonding, we can get a sLMO-HF wavefunction whose quality is very close to the HF
wavefunction.

\begin{table*}
  \centering
  \begin{tabular*}{\textwidth}{@{\extracolsep{\fill}}l r r r r r r@{}}
  Basis set & E(Prism) & Book & Cage &
  Open & Prb & Chair \\
  \hline
  Minimal & & & & & & \\
  \hline
  cc-pVDZ & -456.130 & +1.786 & +3.741 & +3.231 & +3.272 & +0.292 \\
  cc-pVTZ & -456.336 & +3.148 & +3.571 & +4.732 & +4.128 & +3.195 \\
  cc-pVQZ & -456.391 & +3.844 & +3.345 & +5.284 & +4.367 & +4.419 \\
  cc-pV5Z & -456.410 & +2.850 & +3.035 & +4.221 & +3.494 & +2.405 \\
  aug-cc-pVDZ & -456.231 & +3.267 &  +2.762  &  +4.602 &  +4.069  & +4.248 \\
  aug-cc-pVTZ & -456.369 & +2.140 &  +2.855  &  +3.554 &  +3.047  & +1.286 \\
  aug-cc-pVQZ & -456.411 & +1.722 &  +2.747  &  +3.096 &  +2.665  & +0.666 \\
  aug-cc-pV5Z & -456.424 & +1.165 &  +2.536  &  +2.447 &  +2.225  & -0.171  \\
  Extended & & & & & & \\
  \hline
  cc-pVDZ  & -456.195 & +2.242 & +3.264  & +3.231 &  +2.971 &  +0.771 \\
  cc-pVTZ  & -456.373 & +2.952 & +2.928  & +4.206 &  +3.546 &  +2.896 \\
  cc-pVQZ  & -456.416 & +2.855 & +2.890  & +4.135 &  +3.460 &  +2.795 \\
  cc-pV5Z  & -456.427 & +2.008 & +2.718  & +3.302 &  +2.799 &  +1.179 \\
  aug-cc-pVDZ  & -456.274 & +1.759 & +2.550  & +3.028 &  +2.751 & +1.177 \\
  aug-cc-pVTZ  & -456.390 & +1.622 & +2.617  & +2.962 &  +2.619 & +0.890 \\
  aug-cc-pVQZ  & -456.422 & +1.348 & +2.569  & +2.675 &  +2.402 & +0.231 \\
  aug-cc-pV5Z  & -456.431 & +0.703 & +2.408  & +2.006 &  +1.938 & +0.783 \\
  Hbond     & & & & & & \\
  \hline
  cc-pVDZ & -456.159 & +2.384 & +4.427   & +3.650 & +3.922 &  +1.803 \\
  cc-pVTZ & -456.354 & +0.606 & +3.063   & +1.898 & +2.397 &  -0.528 \\
  cc-pVQZ & -456.410 & +0.372 & +2.323   & +0.873 & +1.492 &  -1.772 \\
  cc-pV5Z & -456.429 & +1.126 & +1.948   & +0.103 & +0.881 &  -2.907 \\
  HF     & & & & & & \\
  \hline
  cc-pVDZ & -456.239 & +4.675 & +4.175 & +5.330 & +4.877 & +6.009 \\
  cc-pVTZ & -456.399 & +0.716 & +2.622 & +1.600 & +2.039 & -0.039 \\
  cc-pVQZ & -456.437 & +0.771 & +2.080 & +0.305 & +1.067 & -2.382 \\
  cc-pV5Z & -456.447 & +1.400 & +1.801 & +0.251 & +0.637 & -3.401 \\
  aug-cc-pVDZ & -456.296 & -0.230 & +2.174 & +0.726 & +1.416 & -1.477 \\
  aug-cc-pVTZ & -456.408 & -1.403 & +1.804 & -0.251 & +0.627 & -3.383 \\
  aug-cc-pVQZ & -456.439 & -1.473 & +1.754 & -0.322 & +0.583 & -3.489 \\
  aug-cc-pV5Z & -456.447 & -1.508 & +1.737 & -0.356 & +0.555 & -3.548 \\
  \end{tabular*}
  \caption{Relative Energies (in mhartree) to Prism Conformer calculated for
    (H\textsubscript{2}O)\textsubscript{6} with various fragmentation patterns and basis
    sets. \label{tab:water_energies}}
\end{table*}

Unfortunately, this strategy of extending the fragments to increase the quality
of the sLMO-HF wavefunction can come with a drawback. We find that when we
extend the fragments on many molecules, convergence becomes significantly more
difficult, which is in agreement with previous observations made by Fornili et
al.\cite{fornili2003} who suggested that the issues could come from stronger
coupling between equations. However, this is not always the case, we use
expanded fragments in both the extended and hbond water calculations. The
average number of iterations to converge across the six conformers are given in
Table \ref{tab:water_iteration}. While we see this increase in convergence
difficulty for the hbond fragmentation, where fragments are shared between
water molecules, we do not see the same problem for the extended fragments,
where each water is a fragment. In fact, the extended fragmentation converges
faster than the minimal fragmentation. This is likely because the extended
fragmentation here is simply each water molecule.

\begin{table}
  \centering
  \begin{tabular}{l r r r}
  basis set & Minimal & Extended & Hbond \\
  \hline
  cc-pVDZ   & 103.17 & 49.17 & 725.50 \\
  cc-pVTZ   & 95.83  & 49.00 & 1069.33 \\
  cc-pVQZ   & 132.17 & 60.67 & 1282.00 \\
  cc-pV5Z   & 185.50 & 72.50 & 1443.33 \\
  \end{tabular}
  \caption{Average number of iterations to converge for the six water conformers.
  \label{tab:water_iteration}}
\end{table}

As mentioned above, one of our primary interests in sLMO-HF is as a method to
generate a wavefunction suitable for post-HF methods. In order for many post-HF
methods, such as MP2, to be able to pick up subtle effects like dispersion, it
is necessary to include many diffuse functions. This can cause difficulties in
convergence due to the increasing number of linearly dependent basis functions.
These difficulties stem from the fact that when there are linear dependencies,
the eigenvalues of the overlap matrix get very small, so when orthogonalizing
the basis, the elements of the Fock matrix can get very large, which can lead
to numerical errors. Of course, this effect can be tempered by removing overlap
eigenvectors with eigenvalues below an appropriately chosen cut-off value, but
this can lead to other issues. For example, it has been pointed out that when
performing a geometry optimization, the eigenvalues of the overlap will change,
so certain basis functions could be included for some geometries but not all,
leading to ambiguity in the definition of the basis set. These problems are
accentuated in large globular structures like peptides and proteins. To this
point, we present in Table \ref{tab:alanine_energies} the sLMO-HF and HF
energies for a peptide consisting of 10 alanine amino acid groups arranged in
an alpha helix. The fragmentation pattern used here restricts lone pair and
core electrons to basis functions on the centers of the corresponding heavy
atoms and bonding electrons to basis functions on the centers of the atoms of
the bond. We ran the HF calculations discarding eigenvectors of the overlap
that have a eigenvalue less than $10^{-5}$ and using a convergence criterion of
$10^{-8}$ change in the root mean square density between iterations. Here we
see that as we include many diffuse functions from the d-aug-cc-pVDZ basis set
that HF no longer converges, even though we are only using only a double
$\zeta$ basis set. In order to get these calculations to converge, it is
necessary to lower the convergence criterion to $10^{-6}$. However, with
sLMO-HF, due to the localization, many basis functions become orthogonal, so
the number of linearly dependent functions are significantly reduced. This
allows the use of larger basis sets with more diffuse functions that will help
facilitate post-HF methods.

\begin{table}
  \centering
  \begin{tabular}{l r r}
  basis set & sLMO-HF & HF \\
  \hline
  cc-pVDZ       & -2664.261813 &  -2666.695570  \\
  aug-cc-pVDZ   & -2665.134051 &  -2666.836046  \\
  d-aug-cc-pVDZ & -2665.208121 &  not converged  \\
  \end{tabular}
  \caption{Energies (in a.u.) calculated for the alanine peptide with various basis sets
    using sLMO-HF, extended sLMO-HF, and HF. \label{tab:alanine_energies}}
\end{table}

\section{Conclusion}
In this paper, we have proposed the use of sLMO-HF as a method to produce
wavefunctions that are well suited to performing post-HF methods. We presented
an alternative quasi-Newton algorithm for finding sLMO-HF orbitals where we have
used Thouless rotations to optimize the variational parameters as opposed to the
SCF procedure. We also showed that sLMO-HF converges exponentially to CBS limit
and that the sLMO-HF CBS limit is different than the HF CBS limit which is due
to sLMO-HF losing some of the fine structure near the nuclei close to a fragment
that is present in HF. In order to regain those features, basis functions on
those nuclei would need to be included. We discussed three effects of using
different fragmentation schemes and basis sets. First, that for compact
fragmentations, sLMO-HF is less sensitive to basis set size because sLMO-HF is
not as susceptible to BSSE. This is because the definition of the fragments
strictly prevents sharing of many basis functions between fragments. We then
showed that by choosing an appropriate fragmentation, the sLMO-HF wavefunction
can get close to the quality of the HF wavefunction. Lastly, we have shown that
sLMO-HF is capable of converging calculations using basis sets with many diffuse
functions, which HF has large difficulties with. This point is important because
in order for many post-HF methods to perform well, the inclusion of many diffuse
functions is critical.

The main drawback of this method is the number of iterations required for
convergence of the sLMO wavefunction. The non-orthogonal nature of the orbitals
leads to numerical difficulties in converging the energy, which becomes worse
as the fragmentation is expanded to include more atoms and larger basis sets.
It is possible that given a better initial guess of the orbitals the number of
iterations could be significantly reduced. We will leave this question to
future research.
\par\noindent\rule{0pt}{2\baselineskip}\par

\FloatBarrier

\bibliography{slmo_paper.bib}
\bibliographystyle{plain}

\end{document}